# Dopant Solubility, and Charge Compensation in La-doped SrSnO$_3$ Films

Tristan Truttmann[1,*], Abhinav Prakash[1], Jin Yue[1], Thomas E. Mates[2], and Bharat Jalan[1,†]

[1]Department of Chemical Engineering and Materials Science, University of Minnesota,

Minneapolis, MN 55455, USA

[2]Materials Department, University of California, Santa Barbara, Santa Barbara, CA 93106, USA

[*]Corresponding author: trutt009@umn.edu

[†]Corresponding author: bjalan@umn.edu

**Abstract**

We investigate lanthanum (La) as an *n*-type dopant in the strain-stabilized tetragonal phase of SrSnO$_3$ grown on GdScO$_3$ (110) using a radical-based hybrid molecular beam epitaxy approach. Fully coherent, epitaxial films with atomically smooth film surface were obtained irrespective of doping density. By combining secondary ion mass spectroscopy and Hall measurements, we demonstrate that each La atom contributes to one electron to the film confirming it occupies Sr-site in SrSnO$_3$ and that it is completely activated. Carrier density exceeding $1 \times 10^{20}$ cm$^{-3}$ was achieved in LSSO films, which is in excellent agreement with the dopant-solubility limit predicted by the density functional theory calculations. A record-high room-temperature mobility of 70 cm$^2$V$^{-1}$s$^{-1}$ at $1 \times 10^{20}$ cm$^{-3}$ was obtained in 12 nm La-doped SrSnO$_3$ film making this the thinnest perovskite oxide semiconductor with a reasonably high electron mobility at room temperature. We discuss the structure-dopant-transport property relationships providing essential knowledge for the design of electronic devices using these materials.

Alkaline-earth stannates, in particular cubic BaSnO$_3$ (BSO), have attracted significant attention for their wide bandgap and excellent dopability with reasonably high room-temperature electron mobilities reaching up to 320 cm$^2$V$^{-1}$s$^{-1}$ in bulk single crystals[1], and 183 cm$^2$V$^{-1}$s$^{-1}$ in thin films.[2] These high mobilities are attributed to low electron effective mass and weak electron-phonon scattering at room temperature.[3,4] The mobility-limiting factor in BSO thin films is ascribed predominantly to the large threading dislocation density from the film/substrate lattice mismatch.[2,3,5,6] Insulating buffer layers have been used in an effort to reduce the density of threading dislocations,[2,3,6,7] but although they have resulted in moderate improvements in mobility, the lack of coherent films still stands between BSO and the goal of bulk-like electronic transport. Some progress has been made with the synthesis of high-quality substrates with larger lattice parameters but this effort is still at its infancy.[8-11] Likewise, homoepitaxial growth of BSO films has shown limited success with mobilities ≤ 100 cm$^2$V$^{-1}$s$^{-1}$.[12]

SrSnO$_3$ (SSO) shares many similarities with BSO including a wide bandgap (4.0 eV – 5.0 eV),[13,14] and a low electron effective mass.[15] Similar to BSO, the conduction band minimum in SSO is derived predominantly from Sn-5$s$ states offering low electron effective mass.[15-17] Bulk SSO has three non-cubic polymorphs: an orthorhombic phase (*Pnma*) at room temperature, an orthorhombic phase with *Imma* space group at 905 K < T < 1062 K, and a tetragonal phase (*I4/mcm*) at T > 1062 K before transitioning to a cubic structure at 1295 K.[18] Most significantly, coherent films of SSO have already been achieved on commercially available substrates.[15] The tetragonal phase of SSO can also be stabilized at room temperature using epitaxial compressive strain, whereas an orthorhombic phase (*Pnma*) stabilizes under tensile strain.[15] Significant progress has also been made with the fundamental understanding of the band structures,[15-17] optical properties,[13,16,19] carrier localization,[20] electron-electron interaction[21], electrostatic control[22], and defect-driven magnetism.[19,23] Early device work using SSO as a channel material has yielded exciting results with the record-high peak transconductance value, 17 mS/mm in a depletion mode La-doped SSO (LSSO) *n*-channel metal–semiconductor field-effect transistor.[24]

There remain, however, several open questions including the optimal choice of dopant site and ion, dopant solubility, activation energy and the relative importance of crystal structure, defects (ionized vs. neutral) and electron-phonon scattering on electronic transport. Density functional theory calculations suggest lower formation energy of compensating acceptor defects in LSSO than in La-doped BSO limiting the maximum achievable electron density to $\sim 1 \times 10^{20}$ cm$^{-3}$ (i.e. La solubility limit) in LSSO[25] but this is yet to be experimentally tested.

Here, we report a systematic study of La-doping in the strain-stabilized tetragonal phase of SSO films grown on GdScO$_3$ (GSO) (110) by radical-based hybrid molecular beam epitaxy (MBE). We show that our experimental results are in reasonably good agreement with the calculated maximum achievable carrier density in LSSO prior to the onset of carrier compensation.[25] Using high-resolution X-ray diffraction, secondary ion mass spectroscopy (SIMS) and Hall measurements, we demonstrate a one-to-one correspondence between La-dopant concentration and the measured electron density in addition to establishing the structure-dopant-transport relationships in tetragonal phase of LSSO films.

LSSO films were grown on GSO (110) using a radical-based hybrid MBE approach.[26,27] This approach employs a chemical precursor – hexamethylditin (HMDT) – for Sn, conventional solid sources for Sr, La (ultra-high purity of > 99.99%), and a radio-frequency (RF) plasma source for oxygen. La was used as an *n*-type dopant. Films were grown using co-deposition in an ultra-high vacuum MBE chamber (EVO-50, Omicron) with a base pressure of 10$^{-10}$ Torr. All films were grown at a substrate temperature of 900 °C (thermocouple temperature) and under a fixed oxygen pressure of 5×10$^{-6}$ Torr supplied using a RF plasma source operating at 250 W. La doping concentration was tuned by varying the temperature of the effusion cell, $T_{La}$ between 1180 °C and 1220 °C. Stoichiometry optimization and the details of the MBE method is described elsewhere.[15,20] Structural characterization was performed using a high-resolution Panalytical X'Pert thin film diffractometer with Cu $K_\alpha$ radiation. Wide-angle X-ray diffraction (WAXRD) 2θ-ω coupled scans were taken to determine phase purity, and the out-of-plane lattice parameters. Film thicknesses were determined using the X-ray Kiessig fringes. Atomic force microscopy (AFM) was used for surface characterization. SIMS profiles were collected with a Cameca IMS 7f-Auto (Cameca, Gennevilliers, France). To calibrate SIMS for La quantification, the count ratios of $^{139}$La:$^{112}$Sn were compared to a La-ion-implanted standard. The standard was implanted with 2×10$^{14}$ cm$^{-2}$ La at 50 keV with an angle 7º away from perpendicular to the sample to avoid channeling. La concentration in the films was analyzed with a 5 keV O$^{2+}$ primary beam which was rastered across a 150 × 150 μm$^2$ sample area. An electron gun was used to minimize sample charging. Electronic transport measurements were performed in a Quantum Design Physical Property Measurement System (PPMS Dynacool) using a Van der Pauw configuration to measure the sheet carrier density, sheet resistance, and carrier mobility. For Hall measurements, magnetic fields were swept between - 9 T and + 9 T. Indium was used as an ohmic contact.

Figure 1 shows AFM images of 12 nm LSSO/2 nm SSO/GSO (110) as a function of $T_{La}$ revealing atomically smooth film surfaces with root mean square (rms) roughness values between 115 – 314 pm. Figure 2a shows the high-resolution X-ray diffraction scans of these films indicating phase-pure, singe crystalline films with Kiessig fringes. The sample structure is illustrated in figure 2c. The presence of Kiessig fringes reveals uniform films with excellent interface abruptness and smooth morphology. These results further reveal an expanded out-of-plane lattice parameter of 4.106 ± 0.002 Å consistent with the strain-stabilized tetragonal SSO polymorph on GSO (110)[15] irrespective of doping density. Figure 2b shows the corresponding rocking curves of these films revealing a narrow and a broad Gaussian component for all doping levels as illustrated in the inset. The values of full-width at half maxima (FWHM) varied between 0.046-0.053º of the narrow component and 0.33-0.60º for the broad component. The two-Gaussian shape is commonly seen in epitaxial films. The narrow component is usually ascribed to a long-range uncorrelated disorder limited by the substrate, whereas the broad component is attributed to the correlated short-range disorder.[28] Generally, the broad component emerges in films after strain relaxation and therefore, has been associated with the disorder from the dislocation defects.[29] We note, however, that 14 nm SSO grown on GSO (110) are fully coherent,[15] and thus the strain relaxation can not be the cause for this correlated short-range disorder. These results, however, suggest the presence of short-range correlated structural disorder in these films, which may also depend on the doping concentration. To this end, as a measure of structural disorder, we define the intensity ratio of two Gaussian peaks, $I_{broad}/I_{narrow}$ as a "*disorder proxy*". Larger disorder proxy means high degree of short-range correlated structural disorder. Figure 2d shows the disorder proxy as a function of $T_{La}$ revealing that the disorder proxy first increases and then decreases with increasing $T_{La}$. Experimental results are shown using circular symbols with a sample number embedded inside the symbol for convenient illustration of the data. In order to investigate the role of structural disorder on the electrical transport, we correlate the disorder proxy with the electronic transport in the discussion that follows below.

Figure 3a shows the three-dimensional carrier densities at room temperature ($n_{3D}$) of 12 nm LSSO/2 nm SSO/GSO (110) as a function of $T_{La}$. With an exception of sample labeled (iii), with increasing $T_{La}$, $n_{3D}$ increases expectedly, reaching to a maximum value of $1.1 \times 10^{20}$ cm$^{-3}$ followed by a decrease for $T_{La} > 1210$ °C suggesting the La solid solubility limit in LSSO and the onset of carrier compensation.[25] No measurable conduction was observed in the films doped at $T_{La} < 1180$ °C for $t \leq 12$ nm. In contrast, however, thicker LSSO films with $t > 12$ nm showed significant conductivity even for lower $T_{La}$ and the same growth rate. These results suggest the important role of surface depletion. We, therefore, note that the reported $n_{3D}$ in this study, which is determined by dividing the measured 2D carrier density by the film thickness of 12 nm, is smaller than the actual dopant density in the film. This interpretation is consistent with our findings from thicker LSSO films (48 nm LSSO/2nmSSO/GSO (110)) doped at $T_{La} = 1150$ °C, which resulted in $n_{3D} = 7.5 \times 10^{19}$ cm$^{-3}$. This density is higher than the value obtained at $T_{La} = 1180$ °C in 12 nm LSSO films confirming significant carrier depletion in thinner films. For a reference, figure S1 shows calculated depletion widths vs. $n_{3D}$ assuming different surface potential.[30] To further investigate whether or not the dopant atoms are fully activated, we performed a SIMS depth-profile measurement on 48 nm LSSO/2 nm SSO/GSO (110), doped at $T_{La} = 1150$ °C. La concentration was quantified using SIMS calibration standard. Figure 3b shows an excellent one-to-one correspondence between the La concentration (red bold line) and the measured $n_{3D}$ (marked by horizontal dashed line) demonstrating fully activated dopants in LSSO films. This result further reveals no measurable surface depletion effect in 48 nm LSSO/2 nm SSO/GSO (110) at given doping density. It is noted that a small discrepancy between the La-dopant concentration and the $n_{3D}$ near the film surface (over ~10 nm) as well as a sharp decay (about 7 nm/decade) of La-signal at the SSO/GSO interface are known SIMS artifacts.

Figure 3c shows the room-temperature mobility ($\mu_{RT}$) of these samples as a function of $T_{La}$. With increasing $T_{La}$, $\mu_{RT}$ first increases, and then decreases followed by an increase for 1185 °C $\leq T_{La} \leq$ 1210 °C reaching a maximum value of 70 cm$^2$V$^{-1}$s$^{-1}$ at $1.1 \times 10^{20}$ cm$^{-3}$. For $T_{La} >$ 1210 °C as marked by a vertical shaded line, $\mu_{RT}$ decreases moderately suggesting increased scattering. The moderate decrease in mobility above 1210 °C is likely related to the solubility limit illustrated in figure 3a through either the creation of defects or the loss of carrier screening from the drop in carrier concentration. However, the increase in mobility with increasing La density for 1185 °C $\leq T_{La} \leq$ 1210 °C is surprising. A similar behavior has been seen in compensated semiconductors and is attributed to the charged-impurity scattering. For instance, in relaxed La-doped BSO films, such behavior is ascribed to scattering by charged dislocations.[3,31] In this context, the observed behavior in LSSO films is surprising given these films are fully coherent[15] and *may* also suggest that the source of these disorder is charged. However, since the disorder proxy decreases with increasing doping, it is difficult to comment on the charge-state of the source of these disorders. To this end, we plotted $\mu_{RT}$ as a function of the disorder proxy in figure 3d revealing a notable dependence of mobility on disorder proxy. It was also found that the lowest doped sample (labeled "i") despite having the lowest disorder proxy resulted in lower mobility. These results suggest that room temperature mobility in LSSO films in the measured doping density range is *largely* limited by the structural disorder. However, the origin of these structural disorders and their doping dependence is yet to be understood and will be a subject of future study.

In summary, we have investigated La-doping in the tetragonal phase of SSO revealing a strong correlation between doping density and structural disorder. The room-temperature carrier mobility of 70 cm$^2$V$^{-1}$s$^{-1}$ at $1.1 \times 10^{20}$ cm$^{-3}$ was obtained in 12 nm LSSO films. A carrier density exceeding $1 \times 10^{20}$ cm$^{-3}$ was observed prior to the onset of carrier compensation suggesting the La solubility limit of the order of $10^{20}$ cm$^{-3}$ in tetragonal phase of SSO grown on GSO (110). Using SIMS and the Hall measurements, we show the La concentration matches one-to-one with the measured $n_{3D}$ suggesting fully activated La dopant atoms. At low doping, surface depletion becomes important and accounts for the loss of conductivity in thin films. Future work should focus on exploring the doping dependence of structural disorder as well as on investigating the origin of surface depletion.

**Acknowledgements**

This work was supported through the Young Investigator Program of the Air Force Office of Scientific Research (AFOSR) through Grant No. FA9550-16-1-0205, and Grant No. FA9550-19-1-0245. Part of this work is supported by the National Science Foundation through DMR-1741801, and partially by the UMN MRSEC program under Award No. DMR-1420013. Parts of this work were carried out in the Characterization Facility, University of Minnesota, which receives partial support from NSF through the MRSEC program. We also acknowledge partial support from the renewable development funds (RDF) of the Institute on the Environment (UMN) and the Norwegian Centennial Chair Program seed funds.

**References**

1. H. J. Kim, U. Kim, H. M. Kim, T. H. Kim, H. S. Mun, B.-G. Jeon, K. T. Hong, W.-J. Lee, C. Ju, K. H. Kim, and K. Char, Appl. Phys. Express **5** (6), 061102 (2012).
2. H. Paik, Z. Chen, E. Lochocki, A. Seidner H, A. Verma, N. Tanen, J. Park, M. Uchida, S. L. Shang, B.-C. Zhou, M. Brützam, R. Uecker, Z.-K. Liu, D. Jena, K. M. Shen, D. A. Muller, and D. G. Schlom, APL Mater. **5** (11), 116107 (2017).
3. A. Prakash, P. Xu, A. Faghaninia, S. Shukla, J. W. Ager III, C. S. Lo, and B. Jalan, Nat. Commun. **8**, 15167 (2017).
4. K. Krishnaswamy, B. Himmetoglu, Y. Kang, A. Janotti, and C. G. Van de Walle, Phys. Rev. B **95** (20), 205202 (2017).
5. Woong-Jhae Lee, Hyung Joon Kim, Jeonghun Kang, Dong Hyun Jang, Tai Hoon Kim, Jeong Hyuk Lee, and Kee Hoon Kim, Ann. Rev. Mater. Research **47**, 391 (2017).
6. S. Raghavan, T. Schumann, H. Kim, J. Y. Zhang, T. A. Cain, and S. Stemmer, APL Mater. **4** (1), 016106 (2016).
7. A.P.N. Tchiomo, W. Braun, B.P. Doyle, W. Sigle, P. van Aken, J. Mannhart, and P. Ngabonziza, APL Mater. **7**, 041119 (2019).
8. R Uecker, R Bertram, M Brutzam, Z Galazka, T. M. Gesing, C Guguschev, D Klimm, M Klupsch, A Kwasniewski, and D. G. Schlom, J. Cryst. Growth **457**, 137 (2017).
9. Dong Hyun Jang, Woong-Jhae Lee, Egon Sohn, Hyung Joon Kim, Dongmin Seo, Ju-Young Park, E. J. Choi, and Kee Hoon Kim, J. Appl. Phys. **121**, 125109 (2017).
10. D. Souptel, G. Behr, and A. M. Balbashov, J. Crys. Growth **236**, 583 (2002).
11. Cong Xin, Philippe Veber, Mael Guennou, Constance Toulouse, Nathalie Valle, Monica Ciomaga Hatnean, Geetha Balakrishnan, Raphael Haumont, Romuald Saint Martin, Matias Velazquez, Alain Maillard, Daniel Rytz, Michael Josse, Mario Maglione, and Jens Kreisel, CrystEngComm **21**, 502 (2019).
12. W.-J. Lee, H. J. Kim, E. Sohn, T. H. Kim, J.-Y. Park, W. Park, H. Jeong, T. Lee, J. H. Kim, K. -Y. Choi, and K. H. Kim, Appl. Phys. Lett **108**, 082105 (2016).
13. T. Schumann, S. Raghavan, K. Ahadi, H. Kim, and S. Stemmer, J. Vac. Sci. Technol. A **34**, 050601 (2016).
14. E. Baba, D. Kan, Y. Yamada, M. Haruta, H. Kurata, Y. Kanemitsu, and Y. Shimakawa, J. Phys. D: Appl. Phys, **48**, 455106 (2015).
15. Tianqi Wang, Abhinav Prakash, Yongqi Dong, Tristan Truttmann, Ashley Bucsek, Richard James, Dillon D. Fong, Jong-Woo Kim, Philip J. Ryan, Hua Zhou, Turan Birol, and Bharat Jalan, ACS Appl. Mater. Interfaces **10**, 43802 (2018).
16. David J. Singh, Qiang Xu, and Khuong P. Ong, Appl. Phys. Lett. **104** (1), 011910 (2014).
17. K. P Ong, X Fan, A Subedi, M. B Sullivan, and D. J. Singh, APL Mater. **3**, 062505 (2015).
18. Marianne Glerup, Kevin S. Knight, and Finn Willy Poulsen, Mater. Res. Bull. **40** (3), 507 (2005).
19. D Gao, X. Gao, Y. Wu, T. Zhang, J. Yang, and X. Li, Physica E Low Dimens Syst Nanostruct. **109**, 101 (2019).
20. T. Wang, L. R. Thoutam, A. Prakash, W. Nunn, G. Haugstad, and B. Jalan, Phys. Rev. Mater. **1** (6), 061601 (2017).

21  Jin Yue, Laxman R. Thoutam, Abhinav Prakash, Tianqi Wang, and Bharat Jalan, arXiv:1905.07810 [cond-mat.mtrl-sci] (2019).
22  Laxman Raju Thoutam, Jin Yue, Abhinav Prakash, Tianqi Wang, Kavinraaj Ella Elangovan, and Bharat Jalan, ACS Appl. Mater. Interfaces **11**, 7666 (2019).
23  Q. Gao, H. L. Chen, K. F. Li, and Q. Z. Liu, ACS Appl. Mater. Interfaces **11**, 18051 (2019).
24  V. R. S. K. Chaganti, A. Prakash, J. Yue, B. Jalan, and S. J. Koester, IEEE Electr. Device L. **39** (9), 1381 (2018).
25  L. Weston, L. Bjaalie, K. Krishnaswamy, and C. G. Van de Walle, Phys. Rev. B **97** (5), 054112 (2018).
26  Abhinav Prakash, John Dewey, Hwanhui Yun, Jong Seok Jeong, K. Andre Mkhoyan, and Bharat Jalan, J. Vac. Sci. Technol. A **33** (6), 060608 (2015).
27  A. Prakash, P. Xu, X. Wu, G. Haugstad, X. Wang, and B. Jalan, J. Mater. Chem. C **5** (23), 5730 (2017).
28  P. F. Miceli and C. J. Palmstrom, Phys. Rev. B **51**, 5506 (1995).
29  T. Wang, K. Ganguly, P. Marshal, P. Xu, and B. Jalan, Appl. Phys. Lett. **103**, 212904 (2013).
30  See supplementary information for depletion width as a function of dopant concentration in La-doped SSO films
31  U. Kim H. Mun, H. Min Kim, C. Park, T. Hoon Kim, H. Joon Kim, K. Hoon Kim, and K. Char, Appl. Phys. Lett. **102**, 252105 (2013).

**Figures (Color Online):**

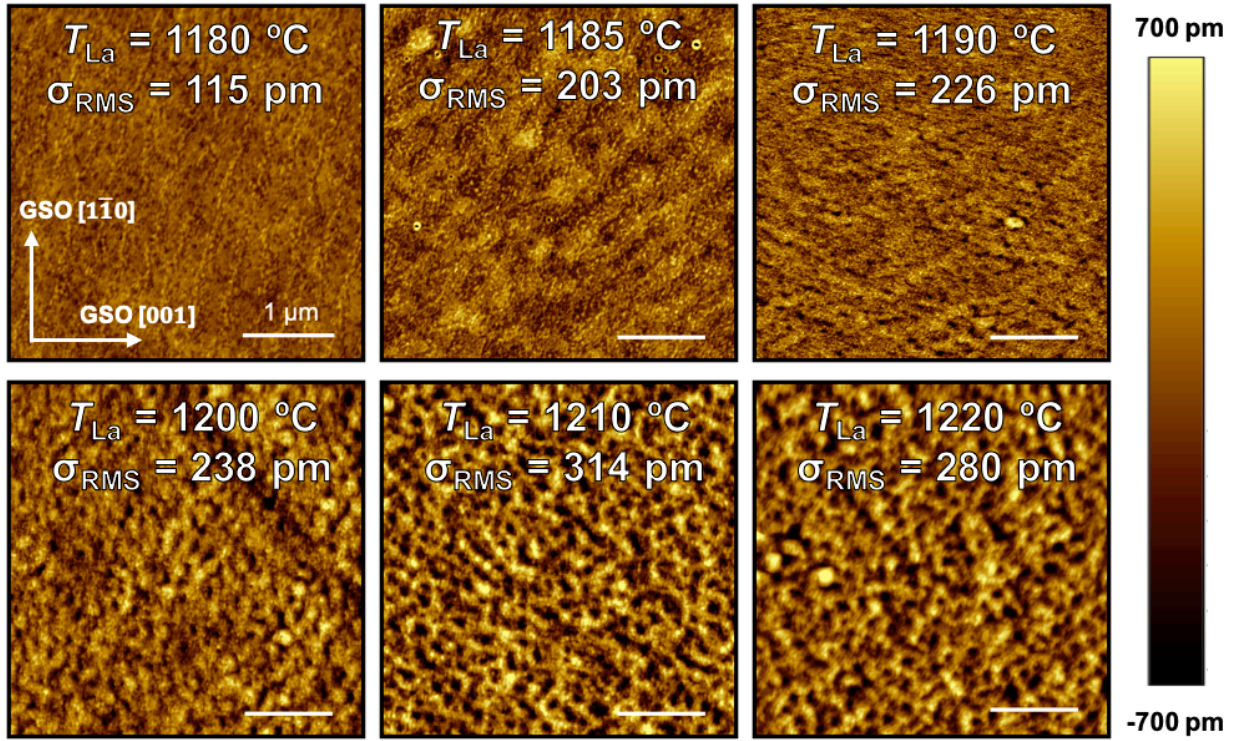

**Figure 1:** AFM images as a function of $T_{La}$ for 12 nm LSSO/2 nm SSO/GSO (110) films showing atomically smooth film surfaces.

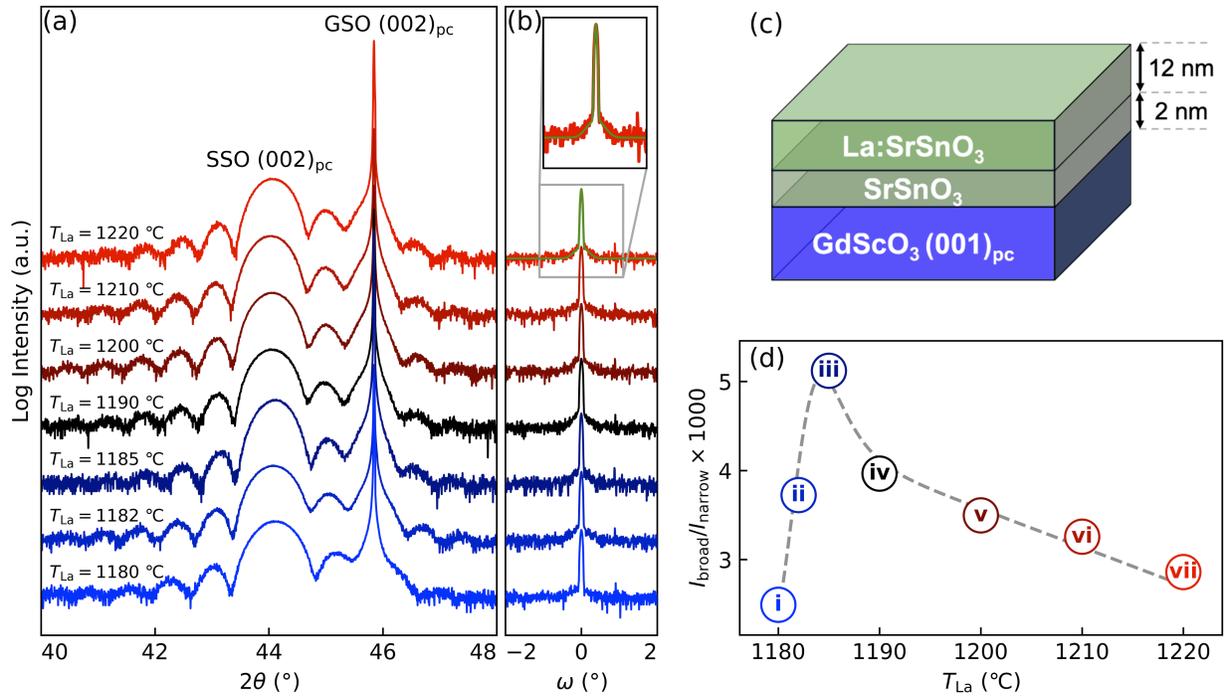

**Figure 2:** (a) On-axis high-resolution X-ray diffraction scans, and (b) the rocking curves around (002) film peak of 12 nm LSSO/2 nm SSO/GSO (110) films as a function of $T_{La}$, (c) A schematic of the sample. The inset of part b shows a narrow and a broad component using a two-Gaussian fit, (d) The intensity ratio of broad and narrow peaks ($I_{broad}/I_{narrow}$) as a function of $T_{La}$.

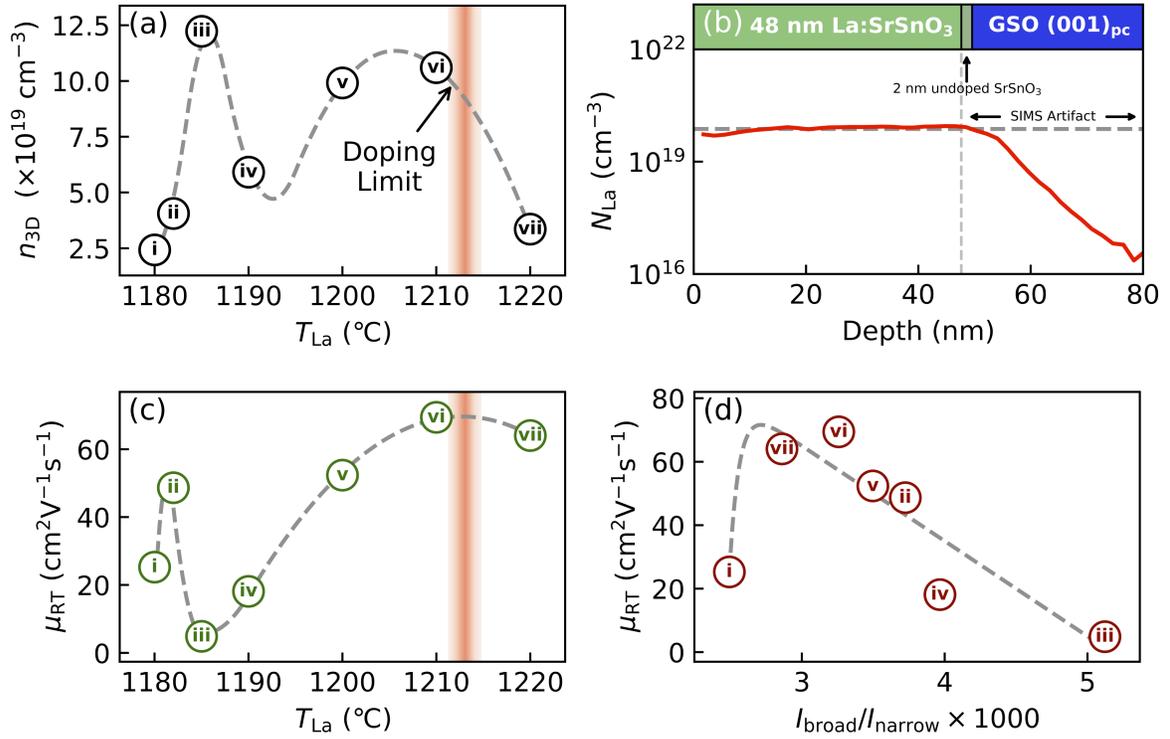

**Figure 3:** (a) $n_{3D}$ as a function of $T_{La}$, (b) SIMS depth-profile showing La-dopant concentration ($N_{La}$) as a function of depth of 48 nm LSSO/2 nm SSO/GSO (110). The horizontal dashed line marks the position corresponding to the measured $n_{3D}$ from Hall measurements indicating one-to-one correspondence with $N_{La}$. $\mu_{RT}$ as a function of $T_{La}$ (c), and $I_{broad}/I_{narrow}$ (d). The vertical line in parts (a) and (c) marks the the onset of charge compensation.

14